\pdfoutput=1
\pdfoutput=1
\pdfoutput=1
\pdfoutput=1
\pdfoutput=1

\documentclass[a4paper,12pt]{report}

\author{Shkerin}
\title{Muonium annihilation}

\usepackage[english]{babel}
\usepackage{amssymb,amsmath}
\usepackage[pdftex]{graphicx}
\usepackage[pdftex]{hyperref}
\usepackage[left=3cm,right=2.5cm,top=2cm,bottom=3cm,bindingoffset=0cm]{geometry}

\begin{document}
\renewcommand\bibname{\Large References}

\begin{titlepage}
\newpage
\begin{center}
\vspace*{6cm}
\LARGE Muonium annihilation into $\nu_e\overline{\nu}_{\mu}$ and $\nu_e\overline{\nu}_{\mu}\gamma$\\[0.5cm]
\large A. Shkerin,\\[0.5cm]
\textit{Institute for Nuclear Research of the Russian Academy of Sciences,}\\
\textit{60th October Anniversary Prospect, 7a, 117312 Moscow, Russia,}\\[0.5cm]
\textit{Moscow Institute of Physics and Technology,}\\
\textit{Institutskii per., 9, 141700, Dolgoprudny, Moscow Region, Russia,}\\[0.5cm]
\textit{E-mail: shkerin@inr.ru}\\[1cm]
\begin{minipage}{0.8\textwidth}
\begin{center}
\textbf{Abstract}
\end{center}
\begin{flushleft}
\large
We calculate in detail the annihilation of Muonium ($Mu$) into $\nu_e\overline{\nu}_{\mu}$ and $\nu_e\overline{\nu}_{\mu}\gamma$ states. For $Mu\rightarrow\nu_e\overline{\nu}_{\mu}$ we obtain the branching ratio $Br=6.6\cdot 10^{-12}$ and for $Mu\rightarrow\nu_e\overline{\nu}_{\mu}\gamma$ in the limit of high energy photon $Br=4.3\cdot 10^{-11}$.
\end{flushleft}
\end{minipage}

\end{center}
\end{titlepage}

\subsection*{Introduction}
In the light of new possibilities of experimental verification of the Standard Model one needs to reconcile the different theoretical predictions. The difference has place, e.g., in the case of Muonium ($Mu$) system decay. The dominant decay channel occurs by the muon beta decay: $\mu^+\rightarrow e^+\nu_e\overline{\nu}_{\mu}$, but there is another possible channel (Muonium annihilation): $Mu\rightarrow\nu_e\overline{\nu}_{\mu}$. The corresponding decay rate is rather small but detectable in the planned experiments \cite{Gninenko:2012nt} as an invisible decay of Muonium. The branching ratio of this process was calculated in some detail and was found to be $\sim 10^{-12}$ in the works \cite{MuoniumDecay}, \cite{Chin} and $\sim 10^{-10}$ in the work \cite{India}. To recheck these results we provide here the full calculations and also estimate the full decay width and a photon energy spectrum of the reaction $Mu\rightarrow\nu_e\overline{\nu}_{\mu}\gamma$.
\subsection*{1. $Mu\rightarrow\nu_e\overline{\nu}_{\mu}$ annihilation}
\subsubsection{$\mu^{+}e^{-}\rightarrow\nu_e\overline{\nu}_{\mu}$ process}
We start with a calculation of the amplitude of $\mu^{+}e^{-}\rightarrow\nu_e\overline{\nu}_{\mu}$ process assuming both $\mu^{+}$ and $e^{-}$ to be free states. In our case particles have small energies in comparison with the masses of the weak bosons, so the amplitude has the form
\begin{equation*}
M_0=\frac{G_F}{\sqrt{2}}j^{\mu}_1j_{\mu 2},
\end{equation*}
where
\begin{eqnarray*}
j^{\mu}_1=\overline{v}^s_{\mu}(p')O^{\mu}v^{s'}_{\nu_{\mu}}(k'),\\
j_{\mu 2}=\overline{u}^r_{\nu_e}(k)O_{\mu}u^{r'}_e(p).
\end{eqnarray*}
Here $u$ and $v$ are the solutions to the Dirac equation with positive and negative frequencies respectively, $p,p',k,k'$ are 4-momenta of $e$, $\mu$, $\nu_e$ and $\nu_{\mu}$ respectively, $O^{\mu}=\gamma^{\mu}\left(\gamma^5+1\right)$. For the square of this matrix element neglecting the neutrino masses we have,
\begin{equation*}
\vert M_0\vert^2 = 128G_F^2(k\cdot p')(p\cdot k').
\end{equation*}
Since electron and muon are supposed to be non-relativistic and both neutrinos are ultra relativistic then
\begin{eqnarray}\label{momentums}
p=(m_e,\overrightarrow{p}), & k=(m_{\mu}/2, m_{\mu}/2),\\
p'=(m_{\mu},-\overrightarrow{p}), & k'=(m_{\mu}/2, -m_{\mu}/2),
\end{eqnarray}
so, the final expression for the matrix element squared is

\begin{equation*}
\vert M_0\vert^2 \approx 32G_F^2\left(m_{\mu}^3m_e+m_{\mu}^2\vert\overrightarrow{p}\vert\cos\theta\left(m_{\mu}+\vert\overrightarrow{p}\vert\cos\theta\right)\right),
\end{equation*}
where $\theta$ is the angle between muon $\mu$ and electron neutrino $\nu_e$. Hereafter we will use the leading part of this expression assuming that $\vert\overrightarrow{p}\vert\approx 0$  and $m_e\ll m_{\mu}$, so, the accuracy of our calculations has order $m_e/m_{\mu}$.

\subsubsection{Bound states}

The amplitude associated with the bound state expresses in terms of the amplitude of the free process as

\begin{equation*}
M=\sqrt{2m}\int\dfrac{d^3q}{(2\pi)^3}\widehat{\psi}^*(\overrightarrow{q})\dfrac{1}{\sqrt{2m_{\mu}}}\dfrac{1}{\sqrt{2m_e}}M_0,
\end{equation*}
where $m$ is the mass of the bound state which in our case equals to $m_{\mu}$ within our accuracy, $\widehat{\psi}(\overrightarrow{q)}$ is a Fourier transform of the Schroedinger wave function of the bound state which we set to be equal to the S ground state. Hence, the full angular momentum of the system is determined by the summary spin of $\mu^+$ and $e^-$. Multipliers ($1/\sqrt{2m_e}$), ($1/\sqrt{2m_{\mu}}$) provide the integral normalizing to unity and the factor $\sqrt{2m}$ in front of the integral needs for accordance with the cross section formula \cite{peskin1995introduction}. Since $M_0$ does not contain any dependency on $\overrightarrow{q}$, this expression simplifies, so for its square we have,

\begin{equation*}
\vert M\vert ^2=\dfrac{1}{2m_e}\vert M_0\vert^2\cdot\vert\psi(0)\vert^2.
\end{equation*}
The wave function of the ground state of Muonium is the same as for hydrogen:

\begin{equation*}
\psi(r)=\dfrac{1}{\sqrt{\pi a^3}}e^{-r/a},
\end{equation*}
where $a$ is the Muonium Bohr radius $a^{-1}=\vert\overrightarrow{p}\vert=\alpha m_{rd}$, and $m_{rd}=m_em_{\mu}/(m_e+m_{\mu})\approx m_e$ with our precision. Therefore
\begin{equation}\label{Psi}
\vert\psi(0)\vert^2=\dfrac{m_e^3\alpha^3}{\pi}.
\end{equation}

\subsubsection{Decay width}

Decay width of Muonium is written in a usual way (again $m\approx m_{\mu}$):

\begin{equation*}
\Gamma=\dfrac{1}{2m_{\mu}}\int\vert M\vert^2 d\Pi_2.
\end{equation*}
A two-particle phase volume $d\Pi_2$ for $2\rightarrow 2$ reactions, which are symmetrical with respect to the collision axis, can be reduced to the form

\begin{equation*}
d\Pi_2=\dfrac{1}{16\pi}d\cos\theta\dfrac{2\vert\overrightarrow{k}\vert}{E_{CM}},
\end{equation*}
where $E_{CM}\approx m_{\mu}$ is the energy of particles in the center-mass system and $\vert \overrightarrow{k}\vert$ is the momentum of the outgoing particles, $\vert \overrightarrow{k}\vert\approx m_{\mu}/2$. The fact that the dependence on the polar angle $\theta$ contains only in $\vert M_0\vert^2$ allows us to write,

\begin{equation}\label{Gamma}
\Gamma=\dfrac{1}{4m_em_{\mu}}\vert\psi(0)\vert^2\int\vert M_0\vert^2d\Pi_2.
\end{equation}
Integration over the phase volume and substitution the explicit expression for $\vert\psi(0)\vert^2$ (\ref{Psi}) leads to the expression:

\begin{equation*}
\Gamma=\dfrac{G_F^2\alpha^3m_e^3m_{\mu}^2}{\pi^2}.
\end{equation*}
Averaging over the initial polarizations gives,
\begin{equation*}
\dfrac{G_F^2\alpha^3m_e^3m_{\mu}^2}{4\pi^2}=48\pi\left(\dfrac{\alpha m_e}{m_{\mu}}\right)^3\Gamma_{\mu\rightarrow e\nu_{\mu}\overline{\nu}_e},
\end{equation*}
so the branching ratio is

\begin{equation}\label{Result1}
Br=\Gamma\tau=6.6\cdot 10^{-12}.
\end{equation}

\subsection*{2. $Mu\rightarrow\nu_e\overline{\nu}_{\mu}\gamma$ annihilation}
\begin{figure}[h!]
\renewcommand{\figurename}{Figure}
\begin{minipage}[h]{0.3\linewidth}
\center{\includegraphics[width=1.0\linewidth]{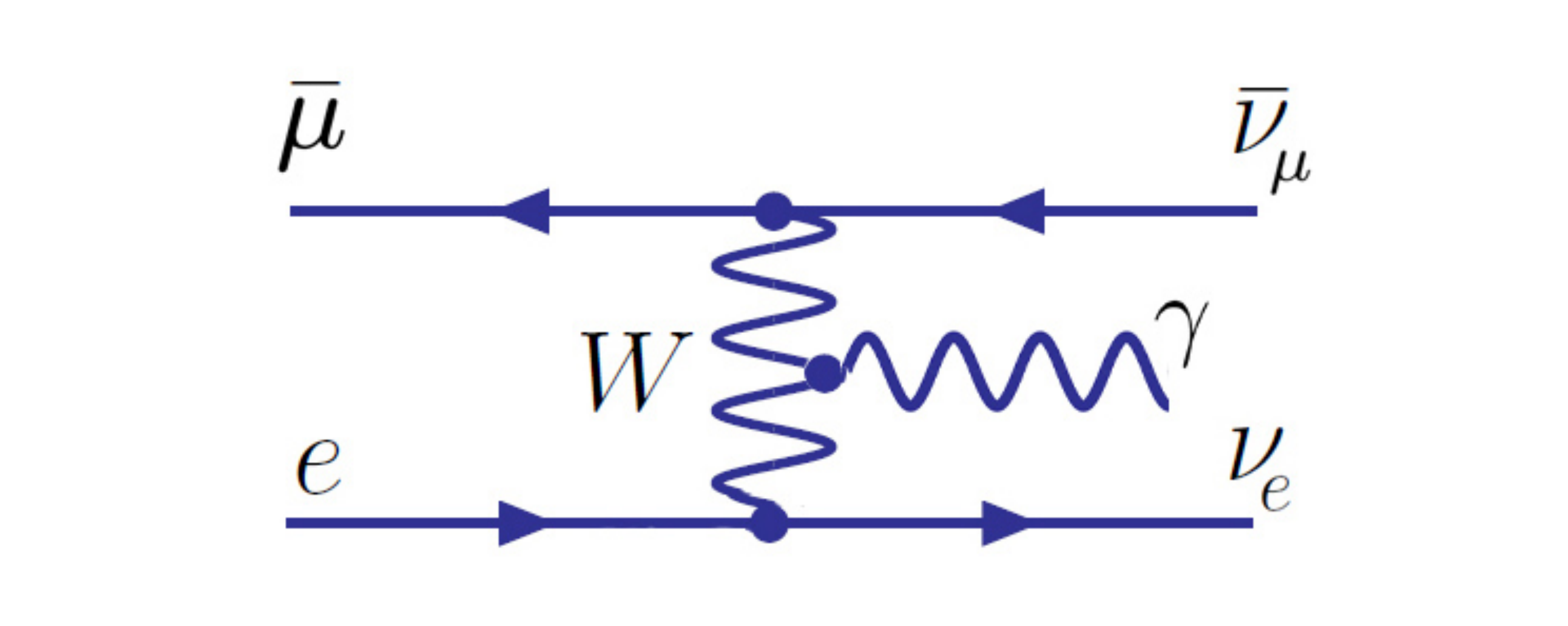}}
\end{minipage}
\begin{minipage}[h]{0.3\linewidth}
\center{\includegraphics[width=1.0\linewidth]{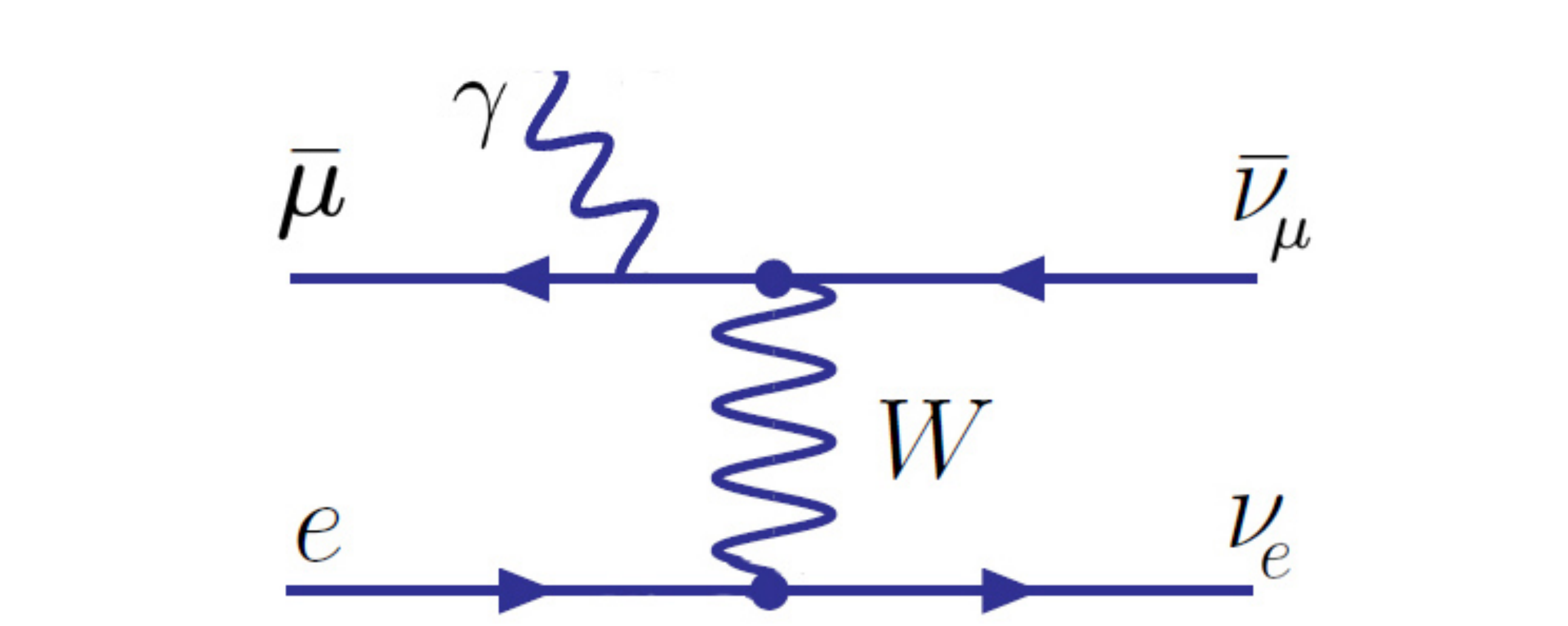}}
\end{minipage}
\begin{minipage}[h]{0.3\linewidth}
\center{\includegraphics[width=1.0\linewidth]{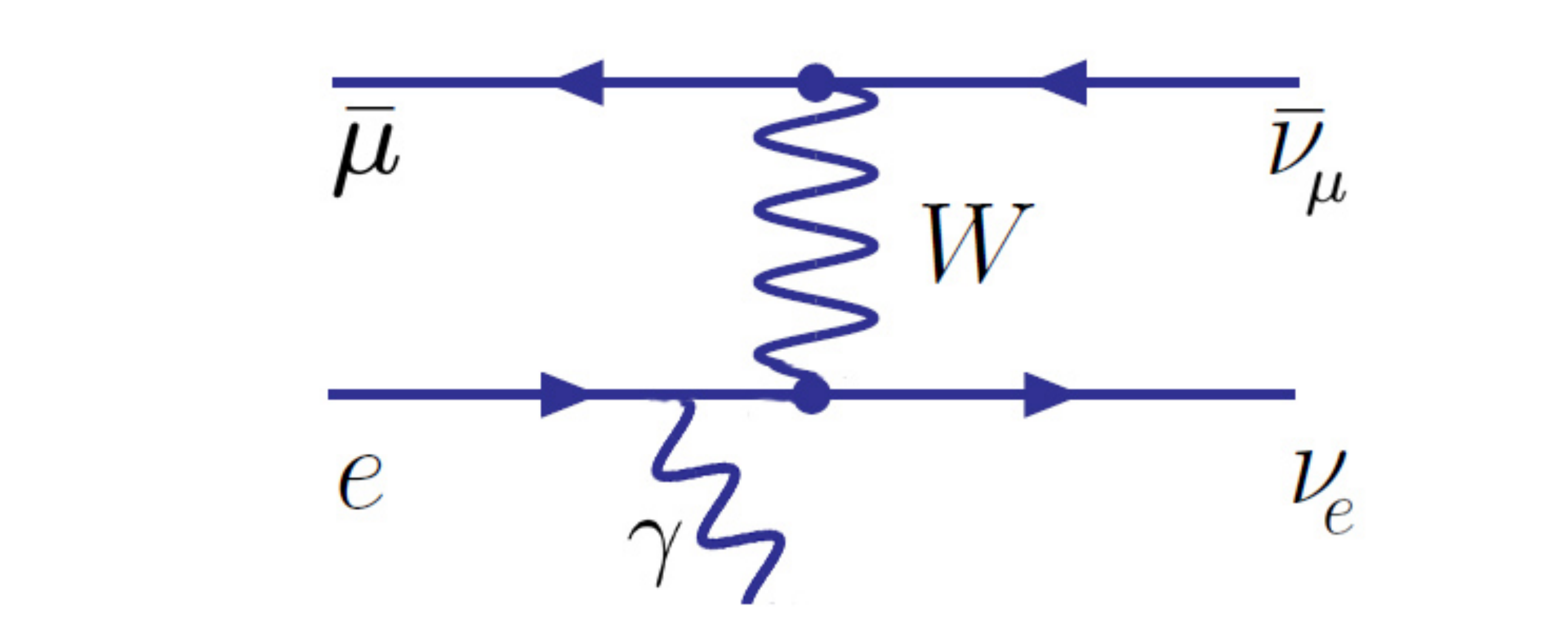}}
\end{minipage}
\caption{The leading order diagrams for free $\mu^+e^-\rightarrow\nu_e\overline{\nu}_{\mu}\gamma$ process.}
\label{Image}
\end{figure}

To the leading order in $\alpha$ there are three diagrams which give contribution to the $Mu\rightarrow\nu_e\overline{\nu}_{\mu}\gamma$ process. They are presented in fig.\ref{Image}.

The first of them contains two virtual W-boson propagators and hence its amplitude is suppressed significantly. Naively, the leading order part of the matrix element squared of $\mu^+e^-\rightarrow\nu_e\overline{\nu}_{\mu}\gamma$ process involving the term from this diagram is proportional to
\begin{equation*}
\vert M_0\vert^2\sim\dfrac{e^2G_F^2}{m_W^2m_e^2},
\end{equation*}
and for its contribution to the full decay width of $Mu$ one can obtain:
\begin{equation*}
\Gamma\sim\dfrac{\alpha^4G_F^2m_{\mu}^7}{m_W^2}\left(1+O\left(\dfrac{m_e}{m_{\mu}}\right)\right).
\end{equation*}
The corresponding branching ratio has an order $Br\sim 10^{-15}$ and turns out to be small even in comparison with the next order loop corrections.
The second diagram is also suppressed by a factor $\sim m_e/m_{\mu}$ and we will not consider it in accordance with the discussion above.
The last diagram describes the emission of a photon from the external electron line. Its amplitude is equal to
\begin{equation*}
\begin{array}{r}
M_{0\gamma}=\dfrac{G_F}{\sqrt{2}}\dfrac{i}{-2p\cdot q}\left(\overline{u}^r_{\nu_e}(k)\gamma_{\mu}(1+\gamma_5)\left(\gamma_{\rho}(p^{\rho}-q^{\rho})-m_e\right)\gamma_{\sigma}\epsilon^{\sigma *}(q)u^s_e(p)\right)\times \\
\times\left(\overline{v}^{s'}_{\mu}(p')\gamma^{\mu}(1+\gamma_5)v^{r'}_{\nu_{\mu}}(k')\right).
\end{array}
\end{equation*}
Here $q$ is the photon momentum. Conjugation, production and averaging over the initial polarizations lead to the following expression,
\begin{equation}\label{SquareGamma}
\vert M_{0\gamma}\vert^2=\dfrac{8e^2G_F^2}{(p\cdot q)^2}\left[2(p\cdot l)(k'\cdot l)(k\cdot p')-l^2(k'\cdot p)(k\cdot p')\right],
\end{equation}
where $l=p-q$. For 4-momenta of the particles we have,
\begin{equation*}
\begin{array}{ccc}
p=(m_e,\overrightarrow{p}), & p'=(m_{\mu},-\overrightarrow{p}), \\
k=(\omega_1, \overrightarrow{k}_1), & k'=(\omega_2, \overrightarrow{k}_2), & q=(\omega_{\gamma}, \overrightarrow{k}_{\gamma}).
\end{array}
\end{equation*} 
For the differential decay width one can write similar to (\ref{Gamma}),
\begin{equation}\label{Gamma2}
\Gamma_{\gamma}=\dfrac{1}{2m_{\mu}}\int\vert M_{\gamma}\vert^2d\Pi_3=\dfrac{\vert\psi(0)\vert^2}{4m_em_{\mu}}\int\vert M_{0\gamma}\vert^2d\Pi_3,
\end{equation}
where $d\Pi_3$ is now a differential three-particle phase volume
\begin{equation*}
\begin{array}{r}
d\Pi_3=(2\pi)^4\delta(\omega_1+\omega_2+\omega_{\gamma}-m_{\mu})\delta(\overrightarrow{k}_1+\overrightarrow{k}_2+\overrightarrow{k}_{\gamma})\dfrac{d^3\overrightarrow{k}_1}{(2\pi)^32\omega_1}\dfrac{d^3\overrightarrow{k}_2}{(2\pi)^32\omega_2}\dfrac{d^3\overrightarrow{k}_{\gamma}}{(2\pi)^32\omega_{\gamma}}.
\end{array}
\end{equation*}
One can integrate over the 3-momenta $\overrightarrow{k}_1$ and $\overrightarrow{k}_2$ using the equality \cite{Okun},

\begin{equation}
\int k_{\alpha}k'_{\beta}\dfrac{d^3\overrightarrow{k}_1}{\omega_1}\dfrac{d^3\overrightarrow{k}_2}{\omega_2}\delta^4(k+k'-Q)=\dfrac{\pi}{6}\left(Q^2g_{\alpha\beta}+2Q_{\alpha}Q_{\beta}\right),
\end{equation}
where $Q=p+p'-q$ in our case. Noting that $d^3\overrightarrow{k}_{\gamma}=4\pi\omega_{\gamma}^2d\omega_{\gamma}$ one can reduce the differential decay width to the form

\begin{equation}\label{DiffDecayRateGamma}
\dfrac{d\Gamma_{\gamma}}{d\omega_{\gamma}}=\dfrac{G_F^2\alpha^4}{12\pi^3}m_em_{\mu}^4F(x),
\end{equation}
where $F(x)=x(3-4x)$ is a photon spectrum function, $x=\omega_{\gamma}/m_{\mu}$, $x\leqslant 0.5$. In accordance with our assumption this expression is valid for $\omega_{\gamma}\gg m_e$, i.e. for $x\gg 5\cdot 10^{-3}$. Integration over $\omega_{\gamma}$ leads to the final expression,
\begin{equation}\label{FreeDecay}
\Gamma_{\gamma}=\dfrac{5G_F^2\alpha^4m_em_{\mu}^4}{288\pi^3}.
\end{equation}
The appropriate branching ratio is $Br=\Gamma_{\gamma}\tau=4.3\cdot 10^{-11}$.

\subsection*{Results}
The branching ratio (\ref{Result1}) for the reaction $Mu\rightarrow\nu_e\overline{\nu}_{\mu}$ is in agreement with the works \cite{MuoniumDecay} and \cite{Chin} but differs from the one obtained in the work \cite{India}. Thus, its value lies in the detectable range of an experiment proposed in \cite{Gninenko:2012nt}. The value (\ref{FreeDecay}) of the full decay width in $Mu\rightarrow\nu_e\overline{\nu}_{\mu}\gamma$ reaction differs slightly from the result in the work \cite{Chin}. The differential decay width (\ref{DiffDecayRateGamma}) is valid for the range of photon energies $\omega_{\gamma}\sim 5-50 MeV$, as we discussed above, and can be also verified in the new experiments.

\subsection*{Acknowledgements}
The author thanks D. S. Gorbunov for useful discussions.

\end{document}